%% file: ms.tex
\documentclass[a4paper]{article}
\usepackage[ninepoint]{itgspeech2021}    %% Include ITGSpeech2021 style
\usepackage{cite}
\usepackage{amsmath,amssymb,amsfonts}
\usepackage{algorithmic}
\usepackage{graphicx}
\usepackage{textcomp}
\usepackage{xcolor}
\def\BibTeX{{\rm B\kern-.05em{\sc i\kern-.025em b}\kern-.08em
    T\kern-.1667em\lower.7ex\hbox{E}\kern-.125emX}}

\usepackage{amsmath,graphicx}

\usepackage{microtype}
\usepackage{times}            %% Choose Times Roman font
\usepackage[english]{babel}   %% This is an English paper
\usepackage[ansinew]{inputenc}%% use 'utf8' or 'ansinew' to support special characters (e.g., umlauts), depends on your editor!
\usepackage[T1]{fontenc}      %% use PS Type1 fonts
\usepackage[sort&compress,numbers]{natbib}	%% bibliography
\usepackage[colorlinks=false,pdfborder={0 0 0}]{hyperref}
\usepackage{units}
\usepackage{import}
\usepackage{balance}

\usepackage[acronym,nohypertypes={acronym,notation}]{glossaries} 
\usepackage{url}

\usepackage{trfsigns} % Include FFT sign o---o
\usepackage{rotating} % rotation
\usepackage{xspace} % Special space

\usepackage{amsmath,amsfonts,amssymb}
\usepackage{array}
\usepackage{bm} % Bold greek letters
\usepackage{bbm} %Bold math symbols
\usepackage{siunitx} % SI units package replaces sistyle

\usepackage{subfig}
\usepackage{graphicx}
\usepackage{psfrag}
\usepackage{xcolor}
\usepackage{tikz}

\usetikzlibrary{calc}
\usetikzlibrary{tikzmark}

\usepackage{longtable}
\usepackage{tabularx}
\usepackage{glossaries}
\usepackage{url}

\usepackage{paralist}

\input{macros.tex}

\input{tikz_macros.tex}

\input{colors.tex}
\input{glossaries.tex}

\begin{document}
\title{A Database for Research on Detection and Enhancement of Speech Transmitted over HF links}

\author{Jens Heitkaemper$^1$, Joerg Schmalenstroeer$^1$, Joerg Ullmann$^1$, Valentin Ion$^2$, Reinhold Haeb-Umbach$^1$}
\address{
$^1$ Department of Communications Engineering, Paderborn University,  Germany \hfil
$^2$ Plath GmbH, Hamburg, Germany\\
Email: \texttt{\{heitkaemper, schmalen, ullmann, haeb\}@nt.uni-paderborn.de}
}

\maketitle

\begin{abstract}
	In this paper we present an open database for the development of detection and  enhancement algorithms of speech transmitted over HF radio channels. 
	It consists of audio samples recorded by various receivers at different locations across Europe, all monitoring the same single-sideband modulated transmission from a base station in Paderborn, Germany. Transmitted and received speech signals are precisely time aligned to offer parallel data for supervised training of deep learning based detection and enhancement algorithms.
	For the task of speech activity detection two exemplary baseline systems are presented, one based on statistical methods employing a multi-stage Wiener filter with  minimum statistics noise floor estimation, and the other relying on a deep learning approach.
\end{abstract}
%
%\begin{IEEEkeywords}
%    database, speech processing, hf links
%\end{IEEEkeywords}
%%

\input{sections/introduction}

\input{sections/database}

\input{sections/sigpro}

\input{sections/dnn}

\input{sections/experiments}

%%%%%%%%%%%%%%%%%%%%%%%%%%%%%%%%%%%%%%%%%%%%%%%%%%%%%%%%%%%%%%%%%%
\section{Summary} \label{SEC:Conclusion}
%%%%%%%%%%%%%%%%%%%%%%%%%%%%%%%%%%%%%%%%%%%%%%%%%%%%%%%%%%%%%%%%%%
In this paper we presented a database of real ham radio recordings with parallel clean and distorted speech data,  offering a novel challenge for speech activity detection, enhancement, and reconstruction.
The database consists of diverse noise conditions with a \gls{SDR} ranging from \SI{-20}{db} to \SI{5}{dB}.
Additionally, two baseline systems for \gls{SAD} are presented.
In future work, we will evaluate speech enhancement and reconstruction  algorithms on this new database.% and work on improved targets for supervised \gls{ASR} training. \inred{Diesen Zusatz versteht man nicht: warum sind die Labels zu schlecht fuer ASR?}

%%%%%%%%%%%%%%%%%%%%%%%%%%%%%%%%%%%%%%%%%%%%%%%%%%%%%%%%%%%%%%%%%%
\section{Acknowledgements}\label{sec:acknowledgements}
%%%%%%%%%%%%%%%%%%%%%%%%%%%%%%%%%%%%%%%%%%%%%%%%%%%%%%%%%%%%%%%%%%
Computational resources were provided by the Paderborn Center for Parallel Computing.
\newpage
\small
\balance
\bibliographystyle{ieeetr}
\bibliography{ms}

\end{document}

%% file: macros.tex
\usepackage{tikz}
\usepackage{pgfplots}
\usepackage{amssymb}
\usetikzlibrary{positioning,chains,calc,shapes.geometric, shadows, shapes.misc, fit, arrows, shapes.callouts}
\usepackage{tabularx, etoolbox, booktabs}
\usepackage{multirow} % Allows rowspan
\usepackage{bbding}
\usepackage{microtype}
\usepackage{pifont}

\usepackage{trfsigns}

%% file: tikz_macros.tex
\usepackage{tikz}
\usepackage{standalone}
\usepackage{pgfplots}

\usetikzlibrary{positioning,chains,calc,shapes.geometric, shadows, shapes.misc, fit, arrows, shapes.callouts}
\usepgfplotslibrary{groupplots}
\pgfplotsset{compat=newest}
\usetikzlibrary{backgrounds}

%% file: colors.tex
\definecolor{yellow}{RGB}{255, 198, 0}
\definecolor{orange}{RGB}{255, 130, 0}
\definecolor{blue}{RGB}{0, 32, 91}
\definecolor{red}{RGB}{198, 53, 39}
\definecolor{magenta}{RGB}{138, 27, 97}
\definecolor{lightblue}{RGB}{0, 159, 223}
\definecolor{green}{RGB}{0, 155,119}
\definecolor{lightgreen}{RGB}{132, 189,0}
\definecolor{greenyellow}{RGB}{208, 223,0}

%% file: glossaries.tex
\newacronym{DNN}{DNN}{deep neural network}
\newacronym{CDR}{CDR}{coherent-to-diffuse power ratio}
\newacronym{DoA}{DoA}{direction-of-arrival}
\newacronym{TDoA}{TDoA}{time difference of arrival}
\newacronym{MSE}{MSE}{mean-squared error}
\newacronym{FF}{FF}{feed forward}
\newacronym{MLP}{MLP}{multilayer perceptron}
\newacronym{CRNN}{CRNN}{convolutional recurrent neural network}
\newacronym{RNN}{RNN}{recurrent neural network}
\newacronym[plural=GPs,firstplural=Gaussian processes (GPs)]{GP}{GP}{Gaussian process}
\newacronym{GRU}{GRU}{gated recurrent unit}
\newacronym{RIR}{RIR}{room impulse response}
\newacronym{AWGN}{AWGN}{additive white Gaussian noise}
\newacronym{SNR}{SNR}{signal-to-noise ratio}
\newacronym{STFT}{STFT}{short-time Fourier transform}
\newacronym[plural=PSDs,firstplural=power spectral densities~(PSDs)]{PSD}{PSD}{power sprectral density}
\newacronym{AE}{AE}{absolute error}
\newacronym{MAE}{MAE}{mean-absolute error}
\newacronym{PE}{PE}{position error}
\newacronym{MPE}{MPE}{mean position error}
\newacronym{WASN}{WASN}{wireless acoustic sensor network}
\newacronym{OoR}{OoR}{out-of-range}
\newacronym{ASR}{ASR}{automatic speech recognition}
\newacronym{TDNN}{TDNN}{time delay neural network}
\newacronym{CNN}{CNN}{convolutional neural network}
\newacronym{WLS}{WLS}{weighted least squares}
\newacronym{LS}{LS}{least squares}
\newacronym{LPC}{LPC}{linear predictive coding}
\newacronym{RANSAC}{RANSAC}{random sample consensus}

\newacronym{GARDE}{GARDE}{Geometry cAlibration fRom Distance Estimates}
\newacronym{MDS}{MDS}{Multi Dimensional Scaling}

\newacronym{GMM}{GMM}{gaussian mixture model}
\newacronym{HMM}{HMM}{hidden Markov model}

\newacronym{CWLS}{CWLS}{constrained weighted least squares}
\newacronym{CRLB}{CRLB}{Cramer-Rao lower bound}
\newacronym{RMSE}{RMSE}{Root Mean Square Error}

\newacronym{ITU}{ITU}{International Telecommunication Union}
\newacronym{SDR}{SDR}{software defined radio}
\newacronym{SAD}{SAD}{speech activity detection}
\newacronym{ROC}{ROC}{receiver operating characteristic}
\newacronym{AUC}{AUC}{area under the curve}
\newacronym{EER}{EER}{equal error rate}
\newacronym{AGC}{AGC}{Automatic Gain Control}
\newacronym{STOI}{STOI}{Short-time Objective Intelligibility}
\newacronym{CSBE}{CSBE}{combined sub-band energy}
\newacronym{SSB}{SSB}{single sideband}
\newacronym{HF}{HF}{high frequency}
\newacronym{RT}{RT}{realtime}

%% file: sections/introduction.tex
%% Tex magic
% !TeX spellcheck = en_US
% !TeX encoding = utf-8
% !TeX root = ../Heit2021.tex
% !TeX program = pdflatex
% !BIB program = bibtex

%%%%%%%%%%%%%%%%%%%%%%%%%%%%%%%%%%%%%%%%%%%%%%%%%%%%%%%%%%%%%%%%%%
\section{Introduction}
%%%%%%%%%%%%%%%%%%%%%%%%%%%%%%%%%%%%%%%%%%%%%%%%%%%%%%%%%%%%%%%%%%
During the last decade \glspl{DNN} have achieved impressive results on a variety of speech tasks, such as  \gls{SAD}, speech enhancement, source separation, and \gls{ASR} \cite{Watanabe18book,Haeb19Overview, Haeb21FarField}. Their success, however, rests crucially on the availability of labeled training data, as the systems are trained in a supervised manner \cite{Heymann18MaskBF}. 

Manual annotation of training data can be a very tedious task, and for some problems, such as speech enhancement, it is nearly impossible for large data sets. This issue can be circumvented with parallel data, which refers to the existence of  distorted and clean versions of the same utterance. While the former is applied to the network input, the latter can be used to automatically derive the training targets. For example, targets for time frequency masks can be generated from clean speech and then used to train a mask estimator from noisy speech \cite{Erdogan15Masks,Kolbaek17UPit}.

Such parallel data is usually generated  by artificially distorting  clean speech signals. Among the many examples of such databases with artificially generated parallel data  are   WSJ0-2Mix \cite{Wang18McDc} and SMS-WSJ \cite{Drude18SMSWSJ}, two data sets for source separation research. 
For real recordings of degraded speech, parallel data is usually not available. For example,  the CHiME-5 \cite{chime5} and AMI \cite{ami} databases offer real recordings  of a meeting scenario,  but parallel clean speech is not available.
%Similarly, two databases offering high frequency transmission data and released in the last decade 
%
%the audio data published recently by the University of Texas to set up the ''Fearless Step challange'' \cite{fearless20} does not offer parallel data.
% The data set (overall 19k hours) of the entire Apollo-11 mission
% is split into several tasks, comprising \gls{SAD}, speaker diarization, speaker identification and \gls{ASR}.
%However, the database does not offer parallel data.
% Similarly, the recently released audio data of the Apollo-11 mission does not offer parallel data \cite{fearless20}.

This leaves researchers with a dilemma: while artificial corruption of clean data offers the opportunity to provide parallel data useful for network training, real data are ``real'', and artificial degradation can never perfectly mimick true recordings of degraded speech \cite{Heymann18MaskBF}. 

%However, \gls{DNN}s rely on large datasets with highly diverse data and dependable target information for supervised training.
%Especially, enhancement and reconstruction rely on parallel data for the highest performance. \cite{Heymann17Beamnet,Higuchi17AdversarialTraining}
%
%There are already a number of published data sets with distorted speech \cite{ami,Barker15Chime3,chime5,Kinoshita16Reverb,Wang18McDc,Drude18SMSWSJ}.
%However, all face the challenge to provide realistic data while at the same time offering reliable targets.
%For speech enhancement the AMI \cite{ami} and CHiME-5 \cite{chime5} data set offer realistic recordings, however no clean parallel data is available.
%In contrast, WSJ-2-Mix \cite{Wang18McDc} an SMS-WSJ \cite{Drude18SMSWSJ} offer perfect target information but only simulated data.
%
%Recently, the audio data of the entire Apollo-11 mission has been published by the University of Texas, setting up the ''Fearless step challange'' \cite{fearless20}. The data set (overall 19k hours) is split into several tasks, comprising \gls{SAD}, speaker diarization, speaker identification and \gls{ASR}.
%Offering realistic data with manual annotated speaker activity information and transcription. However, no parallel data is provided and the \gls{SNR} does not fall below \SI{0}{dB}.

In this paper we present one of the rare cases of a database which offers both, parallel data and real recordings of degraded speech. 
It consists of  \gls{HF} speech transmissions recorded by amateur radio receivers.
Additionally, the original transmitted clean audio, taken from the LibriSpeech database \cite{libri20}, is available and time synchronized with the recordings.
This distinguishes this \gls{HF} database from the two most recently released \gls{HF} databases.
Both the audio data from the entire Apollo-11 mission, which was released by the University of Texas at Dallas during the ''Fearless Step challenge'' \cite{fearless20}
and the \gls{HF} data released as part of the DARPA RATS program \cite{rats12} do not offer parallel data.
% The data set (overall 19k hours) 

Amateur radio (abbrev.\ ''ham radio'') is the non-commercial use of radio frequency spectrum by hobby radio operators. Regulated  by the \gls{ITU} \cite{itu2020}, the service is intended for research purposes, private conversations, and even to provide a means for wireless communication  in emergency or disaster scenarios.

The database presented in this paper can be used to develop speech activity detection, speech enhancement or signal reconstruction algorithms for speech transmitted over HF communication links. This includes not only the aforementioned amateur radio transmissions, but also several commercial applications of the HF radio spectrum, such as aircraft, police and marine radio.

The database has been generated as follows. Our ham radio station located in Paderborn, Germany, transmitted utterances of the Librispeech corpus, which were then received by socalled Kiwi \gls{SDR} stations across Europe \cite{kiwi20}. A WebSocket connection to the Kiwi stations was established to transmit the received signals back to Paderborn via the internet. The received and transmitted data was then synchronized to obtain the desired parallel data sets.

%research purposes, including speech and data driven services. Although ham radio sometimes looks like a technology from the past century compared to the Internet and its data rates, it is also an affordable playground for new technologies and interesting research projects, e.g., data networks like HAMNET \cite{hamnet2020}. The key advantage of ''ham radio'', and more precisely the Kiwi-\gls{SDR} infrastructure, is its diversity of recording conditions and the option to record the same signal at various place in parallel. To the best of our knowledge no database with comparable real audio data in highly adverse noise conditions with parallel clean data has been published so far.

%The fearless data set includes a larger amount of data than our set, but it has much better \gls{SNR} conditions and lacks the diversity of different recording conditions, which can be found in ham radio signal transmissions. The NASA equipment is obviously more sophisticated than a simple antenna configurations and cheap receivers to be found in amateur ham radio. Therefore, ham radio stations deliver much more diverse distortions occurring from local influences, antenna mismatch conditions and weather phenomena, offering more challenging data than any high-end transmission from earth to moon.

The developed database is published under the CC BY 4.0\footnote{\url{http://creativecommons.org/licenses/by/4.0/}} license and can be  used for research on \gls{SAD}, speech enhancement, and \gls{ASR} of speech transmitted over HF radio channels.
In this paper we present two baseline systems for \gls{SAD}: One is based on statistical methods, i.e., Wiener filtering using minimum statistics for noise estimation, and the other utilizes \glspl{DNN} to solve the task.

The paper is organized as follows: In Sec.~\ref{SEC:dataset} the speech data transmission and recording are described, followed by Sec.~\ref{SEC:SAD} explaining the two baseline systems where Sec.~\ref{SEC:SigPro} is dedicated to the statistical approach and Sec.~\ref{SEC:DNN} to the \gls{DNN} system. After discussing the experimental results in Sec.~\ref{SEC:Experiments}, the paper concludes with Sec.~\ref{SEC:Conclusion}.

%% file: sections/database.tex
%% Tex magic
% !TeX spellcheck = en_US
% !TeX encoding = utf-8
% !TeX root = ../Heit2021.tex
% !TeX program = pdflatex
% !BIB program = bibtex

\section{Ham Radio Database}\label{SEC:dataset}
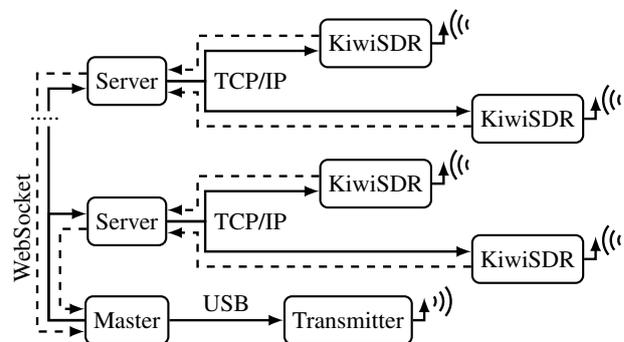
\begin{figure}[htb]
	\input{images/gesamtsystem.tex}
	\caption{System for distributed recording of radio signals.}
	\label{FIG:gesamtsystem}
\end{figure}

To create the database we transmitted speech signals from our amateur radio station in Paderborn and recollected the data which has been received in parallel by several Kiwi-\gls{SDR} stations throughout neighboring countries. Each Kiwi-\gls{SDR} is separately connected to our servers to transmit the recorded signals back to Paderborn via a WebSocket connection.
For the automated transmission the beacon callsign DB0UPB (assigned by the German Telecommunications body ``Bundesnetzagentur'') was used.
The transmission and recording scheme is depicted in Fig.~\ref{FIG:gesamtsystem}.  

Predicted signal quality for the chosen HF carrier frequencies was checked using the web site \cite{VOACAP20}, and appropriate receiver stations from regions with good predicted signal quality were selected. The data set includes recordings from stations in Germany, Austria, Switzerland, Belgium, the Netherlands and the United Kingdom.

The signals are \gls{SSB} modulated using the lower sideband (LSB) with a  bandwidth of $\SI{2.7}{kHz}$ at carrier frequencies of $\SI{7.05}{MHz}-\SI{7.053}{MHz}$ and $\SI{3.6}{MHz} -\SI{3.62}{MHz}$.  
Although, the original audio data has a sampling rate of \SI{16}{kHz}, and the Kiwi-\gls{SDR} sampled the data at \SI{12.001}{Hz} \cite{kiwisoft21}, the finally emitted data is band-limited to $\SI{2.7}{kHz}$, adhering to transmission regulations from the \gls{ITU}. In a first processing step the recorded signals are band-limited to $\SI{4}{kHz}$ via a linear Parks-McClellan filter and downsampled to a rate of $\SI{8}{kHz}$. 
%\inred{Note that the resulting $\SI{8}{kHz}$ signal has a small	sampling rate offset, which will be handled in the post-processing (see Sec.~\ref{SEC:Postprocessing}). RH: ich wuerde den Satz hier weglassen!}

%Removing the upper frequencies significantly improved the \gls{SNR}, because they do not contain speech.
%\inred{Verstehe ich nicht richtig: das ausgesendete SSB-Signal hat doch eine Bandbreite von 2,7kHz, oder?}

\subsection{Data preparation} 
The data preparation targets three key issues: First, the recorded data should allow for automatic annotation of speech activity. Second, the data streams of all recording stations have to be synchronized with the emitted audio data to create parallel data. Aligning the  clean (transmitted) data and noisy (received) data is important to let the speech activity labels computed on the clean carry over to the noisy data. 
Finally, it should be possible to automatically decide whether a transmission has been received  at each individual station, even in very low \gls{SNR} conditions.

\subsubsection{Annotation process}
Our annotation process starts with a carefully conducted data selection and concatenation. The speech samples are taken from the clean training subset of the Librispeech corpus \cite{libri20}, where the subset includes read speech of size close to \SI{500}{hrs} containing over $1000$ speakers.

\begin{figure}[b]
	\centering
	\input{images/example_data}
	\caption{Example of a speech activity  sequence over time. Labels ``sp'' and ``sil'' represent speech and silence, respectively.}
	\label{fig:clean_data}
	%	\caption{.}
\end{figure}
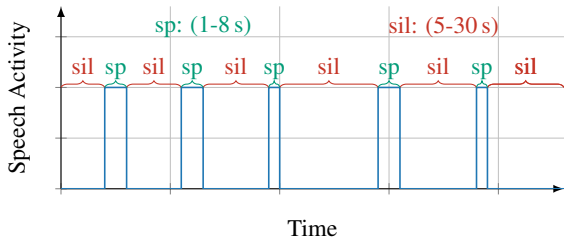

For this database random segments of $\SI{1}{s}$ to $\SI{8}{s}$ length are selected from utterances of the Librispeech corpus.
As shown in Fig.~\ref{fig:clean_data}, we concatenate $5$ segments to a single \textit{audio signal sequence}. Each segment is headed and followed by zeros which guarantee silence periods between utterances. The number of zeros here corresponds to a randomly drawn length between $\SI{8}{s}$ and $\SI{30}{s}$.

Subsequently, the speech activity labels for the audio signal sequences are generated as follows: A \gls{GMM} \gls{HMM} acoustic model is trained on the clean training data set of the Librispeech corpus using the KALDI toolkit~\cite{Povey11Kaldi}. This acoustic model is used to calculate forced alignments,  and all regions with non-silent labels are declared as containing speech.
Additionally, the transcription for each speech segment is generated by extracting the part of the original Librispeech transcription corresponding to the chosen $\SI{1}{s}$ to $\SI{8}{s}$ segment using alignment information.

\subsubsection{Signal preparation for time synchronization}
The Kiwi-\gls{SDR} receivers record a predefined, fixed frequency band and stream the recorded audio data with an unknown time offset via a WebSocket connection to our servers. This unknown offset has to be determined to synchronize the streams from all Kiwi-\glspl{SDR} and align them with the clean transmitted signals.

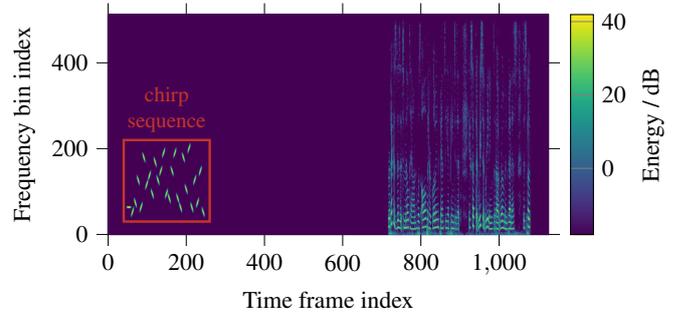
\begin{figure}[b]
	\input{images/marker_signal}
	\caption{Example of a transmitted chirp sequence and the following audio signal}
	\label{FIG:Chirp}
\end{figure}

To ease time synchronization, we added markers before and after each transmitted sequence.
Each marker has a length of \SI{4}{s} and consists of $26$ chirp symbols which differ in starting frequency and orientation (ramp-up or ramp-down). The individual chirp encoding is derived from a gold code to ensure the orthogonality of the markers. In the spectrogram of Fig.~\ref{FIG:Chirp} a chirp sequence can be seen from frame index $0$ to $250$.

Our transmission scheme embeds multiple audio signal sequences
into one single transmission by surrounding the sequences with markers and concatenating $N$ of them.
Additional silence is included after the markers (\SI{5}{s}) and after the audio signal sequences (\SI{1}{s}) to increase the temporal  distance between markers and speech. This is an important aspect to mitigate the effect of the Kiwi-\gls{SDR}'s \gls{AGC}. The \gls{AGC} correctly reacts on the marker and raises the gain, however, in a real audio transmission no preliminary warning is given to the receiver that audio is coming next.
Therefore, the time between the marker and the signal has to be long enough so that the \gls{AGC} readjusts the gain to the original level to get realistic recordings.
The beginning of one transmission scheme is shown in Fig.~\ref{FIG:Chirp}.

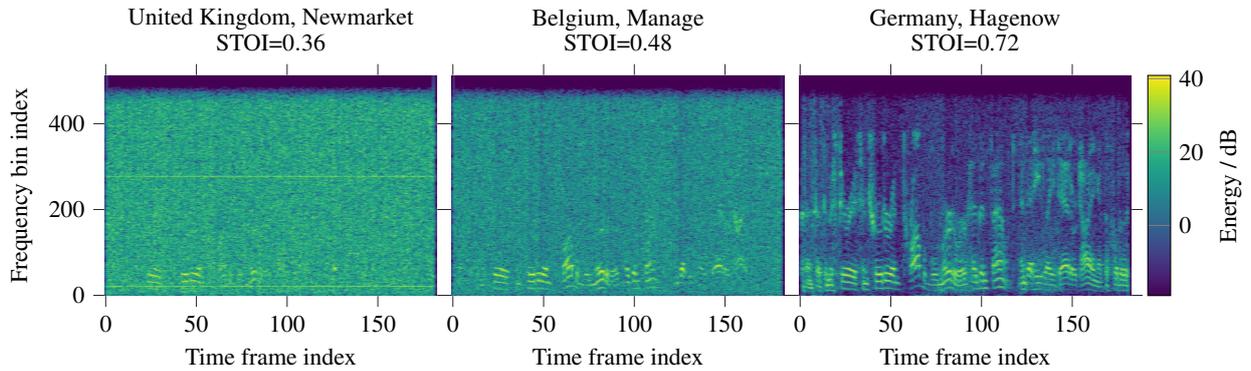
\begin{figure*}[t]
	\centering     %%% not \center
	\input{images/tikz_spect}%
	%	\subfloat[Lausanne, Sw, $\text{STOI} = 0,242$]{\label{FIG:STOI_BAD}\resizebox{0.3 \textwidth}{!}{%
	%			\input{images/stft_17403.tex}
	%	}}
	%	\subfloat[Altenberg, Au, $\text{STOI} = 0,377$]{\label{FIG:STOI_AVG}\resizebox{0.3 \textwidth}{!}{%
	%			\input{images/stft_1809.tex}
	%	}}
	%	\subfloat[Zurich, Sw, $\text{STOI} = 0,667$]{\label{FIG:STOI_GOOD}\resizebox{0.3 \textwidth}{!}{%
	%			\input{images/stft_22012.tex}
	%	}}	
	\caption{Received signals of three  stations recording the same transmitted signal.}
	\label{fig:ex_spects}
\end{figure*}

\subsubsection{Marker based time synchronization}

The recorded Kiwi-\gls{SDR} audio streams are examined in the \gls{STFT} domain. Here, the marker's \gls{STFT} acts as a binary mask which is shifted along the audio stream to find the temporal shift with maximum correlation. Beneficially, the markers are orthogonal to each other and the length of $\SI{4}{s}$ enables a reliable detection even in low \gls{SNR} conditions.

The marker detection process delivers a set of hypotheses for marker positions in the stream. A stream is segmented according to the marker positions and the data is considered valid if the following sanity checks are fulfilled:
\begin{compactitem}
	\item All markers of a transmission are detected.
	\item All markers are detected in the expected order.
	\item All time differences between markers exactly match the expected timing.
\end{compactitem}
The third condition ensures that audio streams with dropped samples are discarded and that the synchronization error cannot exceed \SI{16}{ms}.

Fig.\ref{fig:ex_spects} shows synchronized recordings of the same transmitted audio sample received by three different Kiwi-\gls{SDR} stations.
While some stations deliver audio signals with only few distortions, e.g., Hagenow in Fig.~\ref{fig:ex_spects}, others suffer from bad propagation conditions resulting in low \gls{STOI} measures \cite{Taal2011STOI}, e.g., Newmarket.

%
%\begin{figure*}[htb]
%	\centering
%	\includegraphics[width=1.0\textwidth]{images/korr_overview_good.pdf}
%	\caption{Chirp sequence and audio signal}
%	\label{FIG:Chirp}
%\end{figure*}

\subsubsection{Post processing} \label{SEC:Postprocessing}
The recorded signal has a small sampling rate offset due to the assumption that a \SI{12}{kHz} is recorded which is a \SI{1}{Hz} deviation from the true sampling rate. We correct this sampling rate offset of \SI{56}{ppm} using the approach described in \cite{SroComp18Schmalen} as a post processing step after downsampling.

Furthermore, the marker based synchronization is improved by estimating the delay between the clean signal and the recorded signal and corrected if necessary. Therefore, we estimate the delay of each segment with speech activity using GCC-Phat \cite{GccPhat76Knapp} and calculate the median over all delays in a recorded sequence. If the estimated delay is greater than \SI{16}{ms}, indicating a delay much greater than would be expected from marker-based synchronization, an error due to low \gls{SNR} values is assumed and the signals are left unchanged.

\subsection{Concurrent speakers}
\label{sec:concurrent}
%Ham radio is an open radio frequency used by people from different cultures, mostly following the ham spirit codex. 
Some of the recordings are accidentally corrupted by co-/adjacent channel interferences caused by transmissions from other ham radio users. Some users did not understand the English content explaining the transmission purpose and asked our automatic beacon to leave the frequency. Others did not check the frequency carefully enough and selected transmission frequencies too close to ours. Their signals are transmitted at neighboring frequencies, and, depending on the proximity, are  hardly understandable.
Their speech poses a unique challenge to speech enhancement and reconstruction.
In the data these regions are, obviously, not marked as containing speech, as they have not been transmitted by us. As a consequence a few percent of the labels may not reflect true absence of speech.
% As a consequence, the false alarm rate of \gls{SAD} system may never reach zero due to these ``transcription errors''.

\subsection{Download} \label{SEC:Download}
The database can be downloaded from \url{https://zenodo.org/record/4247491} and code examples on how to train and evaluate a neural network for \gls{SAD} on the presented data can be found in \url{https://github.com/fgnt/ham_radio}. The database is published under the CC BY 4.0 license and is thereby free to use for any purpose, be it commercial or non-commercial.

%% file: images/gesamtsystem.tex
	\tikzset{%
		block/.style    = {draw, thick, rectangle, minimum height = 2em, minimum width = 2em, fill=white, align=center, rounded corners=0.1cm},
		sum/.style      = {draw, circle, node distance = 2cm}, % Adder
		cross/.style={path picture={\draw[black](path picture bounding box.south east) -- (path picture bounding box.north west)
				(path picture bounding box.south west) -- (path picture bounding box.north east);}},
		zigzag/.style = {% added for solution
			to path={ -- ($(\tikztostart)!.55!-9:(\tikztotarget)$) --
				($(\tikztostart)!.45!+9:(\tikztotarget)$) -- (\tikztotarget)
				\tikztonodes},sharp corners}
	}
\begin{tikzpicture}[auto, line width=0.1em]
\node[block, at={(0,0)}] (master) {Master} ;
\node[block, right = of master, xshift=0.5cm] (trans) {Transmitter};
\node[block, above = of master, yshift=-0.35cm] (ssh1) {Server} ;
\node[block, above = of ssh1, yshift=0.2cm] (ssh2) {Server} ;
\node[block, right = of ssh1, yshift=0.5cm, xshift=1cm] (kiwi1) {KiwiSDR} ;
\node[block, right = of ssh1, yshift=-0.5cm, xshift=3cm] (kiwi2) {KiwiSDR} ;
\node[block, right = of ssh2, yshift=0.5cm, xshift=1cm] (kiwi3) {KiwiSDR} ;
\node[block, right = of ssh2, yshift=-0.5cm, xshift=3cm] (kiwi4) {KiwiSDR} ;

\draw[-latex] (master) -- node[above]{USB}(trans);
\draw[-] (master.west) -| node[left, rotate=90, pos=0.9, yshift=0.35cm]{WebSocket} ($(ssh2.west) + (-0.5,-0.6)$);
\draw[-latex] ($(ssh2.west) + (-0.5,-0.4)$) |- ($(ssh2.west) + (0,-0.1)$);
\draw[-, dashed] ($(ssh2.west) + (0, 0.1)$) -| ($(ssh2.west) + (-0.65, -0.4)$);
\draw[-latex, dashed] ($(ssh2.west) + (-0.65,-0.6)$)  |-  ($(master.west) + (0,-0.15)$);
\node[at={($(ssh2.west) + (-0.55,-0.5)$)}] {.....};
\draw[-latex, dashed] ($(ssh1.west) + (0, -0.1)$) -- ($(ssh1.west) + (-0.35,-0.1)$)  |- ($(master.west) + (0,0.15)$);
\draw[-latex] (master.west) -| ($(ssh1.west) + (-0.5,0.1)$) -- ($(ssh1.west) + (0,0.1)$);

\draw[-latex] (ssh1.east) -- ($(ssh1.east) + (0.5,0)$) |- ($(kiwi1.west) + (0,-0.1)$);
\draw[-latex, dashed] ($(kiwi1.west) + (0,0.1)$) -| ($(ssh1.east) + (0.4,0.15)$) --  ($(ssh1.east) + (0,0.15)$);
\draw[-latex] (ssh1.east) -- node[right, pos=1]{TCP/IP} ($(ssh1.east) + (0.5,0)$) |- ($(kiwi2.west) + (0,0.1)$);
\draw[-latex, dashed] ($(kiwi2.west) + (0, -0.1)$) -| ($(ssh1.east) + (0.4,-0.15)$) --  ($(ssh1.east) + (0,-0.15)$);

\draw[-latex] (ssh2.east) -- node[right, pos=1]{TCP/IP} ($(ssh2.east) + (0.5,0)$) |- ($(kiwi3.west) + (0,-0.1)$);
\draw[-latex, dashed] ($(kiwi3.west) + (0,+0.1)$) -| ($(ssh2.east) + (0.4,0.15)$) --  ($(ssh2.east) + (0,0.15)$);
\draw[-latex] (ssh2.east) -- ($(ssh2.east) + (0.5,0)$) |-($(kiwi4.west) + (0,+0.1)$);
\draw[-latex, dashed] ($(kiwi4.west) + (0,-0.1)$) -| ($(ssh2.east) + (0.4,-0.15)$) --  ($(ssh2.east) + (0,-0.15)$);

\draw[-latex] (trans.east) -| ($ (trans.east) + (0.15,0.3) $);
\foreach \r in {.1,.2,.3} \draw ($ (trans.east) + (0.2,0.25) $) ++ (40:\r) arc (40:-40:\r);
\draw[-latex] (kiwi1.east) -| ($ (kiwi1.east) + (0.15,0.3) $);
\foreach \r in {-.1,-.2,-.3} \draw ($ (kiwi1.east) + (0.55,0.25) $) ++ (40:\r) arc (40:-40:\r);
\draw[-latex] (kiwi2.east) -| ($ (kiwi2.east) + (0.15,0.3) $);
\foreach \r in {-.1,-.2,-.3} \draw ($ (kiwi2.east) + (0.55,0.25) $) ++ (40:\r) arc (40:-40:\r);
\draw[-latex] (kiwi3.east) -| ($ (kiwi3.east) + (0.15,0.3) $);
\foreach \r in {-.1,-.2,-.3} \draw ($ (kiwi3.east) + (0.55,0.25) $) ++ (40:\r) arc (40:-40:\r);
\draw[-latex] (kiwi4.east) -| ($ (kiwi4.east) + (0.15,0.3) $);
\foreach \r in {-.1,-.2,-.3} \draw ($ (kiwi4.east) + (0.55,0.25) $) ++ (40:\r) arc (40:-40:\r);
\end{tikzpicture}

%% file: images/example_data.tex
	\tikzset{%
		block/.style    = {draw, thick, rectangle, minimum height = 3em, minimum width = 4em, fill=white, align=center},
		sum/.style      = {draw, circle, node distance = 2cm}, % Adder
		cross/.style={path picture={\draw[black](path picture bounding box.south east) -- (path picture bounding box.north west)
				(path picture bounding box.south west) -- (path picture bounding box.north east);}},
		zigzag/.style = {% added for solution
			to path={ -- ($(\tikztostart)!.55!-9:(\tikztotarget)$) --
				($(\tikztostart)!.45!+9:(\tikztotarget)$) -- (\tikztotarget)
				\tikztonodes},sharp corners}
		%wireless/.pic={
		%	\draw [->] (0,0) -| (.5,#1);
		%	\foreach \r in {.1,.2,.3}
		%	\draw (.6,#1) ++ (60:\r) arc (60:-60:\r);
		%},
		%   connect/.style args={(#1) to (#2) over (#3) by #4}{
		%         insert path={
		%             let \p1=($(#1)-(#3)$), \n1={veclen(\x1,\y1)}, 
		%             \n2={atan2(\y1,\x1)}, \n3={abs(#4)}, \n4={#4>0 ?180:-180}  in 
		%             (#1) -- ($(#1)!\n1-\n3!(#3)$) 
		%             arc (\n2:\n2+\n4:\n3) -- (#2)
		%         }
		%     },
	}
	\tikzstyle{branch}=[{circle,inner sep=0pt,minimum size=0.3em,fill=black}]
	\tikzstyle{box} = [draw, dotted, inner xsep=4mm, inner ysep=3mm]
	\tikzstyle{text_shift} = [xshift=-2em, yshift=0.5em, align=center, font=\footnotesize]
	\definecolor{color0}{rgb}{0.12156862745098,0.466666666666667,0.705882352941177}
	\begin{tikzpicture}[auto, line width=0.1em, node distance = 1cm]
		\begin{axis}[
		height=4cm,
		xticklabels = {},
		yticklabels = {},
		grid=both,
		grid style={line width=.1pt, draw=gray!10},
		major grid style={line width=.2pt,draw=gray!50},
		width=\columnwidth,
		axis lines=left,
		xlabel near ticks,
		ylabel near ticks,
		xmin=0, xmax=46,
		axis line style={-latex},
		ymin=0, ymax=1.8,
		xlabel={Time},
		ylabel={Speech Activity},
%		symbolic x coords={0, 4, 6, 11, 13, 19,20, 29, 30, 31, 38, 39, 46},
%		xticklabels={23,24,25,6,7,8,9,11,12,13,14,15,16,26,27, 17,18,19,20,21,22},
%		xtick=data,
%		/pgf/number format/fixed, 
%		/pgf/number format/precision=2,
%		xticklabel style={name=T\ticknum}% names every xtick label node T0,T1, ...
		]
		\addplot [semithick, solid, color0]
		table {%
			0 0
			4 0
			4 1
			6 1
			6 0
			11 0 
			11 1
			13 1
			13 0
			19 0
			19 1
			20 1 
			20 0
			29 0
			29 1
			31 1
			31 0
			38 0
			38 1
			39 1
			39 0
			46 0
		};
		\begin{scope}[decoration=brace]
		\pgfdecorationsegmentamplitude=5pt
		\draw[decorate, color=red]  (axis cs:0,1) --  (axis cs:4,1) node[above, midway] {sil$\vphantom{p}$};
		\draw[decorate, color=red]  (axis cs:6,1) --  (axis cs:11,1) node[above, midway] {sil$\vphantom{p}$};
		\draw[decorate, color=red]  (axis cs:13,1) --  (axis cs:19,1) node[above, midway] {sil$\vphantom{p}$};
		\draw[decorate, color=red]  (axis cs:20,1) --  (axis cs:29,1) node[above, midway] {sil$\vphantom{p}$};
		\draw[decorate, color=red]  (axis cs:31,1) --  (axis cs:38,1) node[above, midway] {sil$\vphantom{p}$};
		\draw[decorate, color=red]  (axis cs:39,1) --  (axis cs:46,1) node[above, midway] {sil$\vphantom{p}$};
		\draw[decorate, color=red]  (axis cs:39,1) --  (axis cs:46,1) node[above, midway] {sil$\vphantom{p}$};
		\draw[decorate, color=green]  (axis cs:4,1) --  (axis cs:6,1) node[above, midway] {sp};
		\draw[decorate, color=green]  (axis cs:11,1) --  (axis cs:13,1) node[above, midway] {sp};
		\draw[decorate, color=green]  (axis cs:19,1) --  (axis cs:20,1) node[above, midway] {sp};
		\draw[decorate, color=green]  (axis cs:29,1) --  (axis cs:31,1) node[above, midway] {sp};
		\draw[decorate, color=green]  (axis cs:38,1) --  (axis cs:39,1) node[above, midway] {sp};
		\draw (axis cs:13,1.6) node[green](sp){sp: ($1$-\SI{8}{s})};
		\draw (axis cs:35,1.6)  node[red](sil){sil: ($5$-\SI{30}{s})};
		\end{scope}
		\end{axis}
	\end{tikzpicture}

%% file: images/marker_signal.tex
% This file was created by tikzplotlib v0.8.1.
\begin{tikzpicture}

\begin{axis}[
colorbar,
colorbar style={ylabel={Energy / dB}, at={(1.05,1)}, width=0.3cm},
colormap/viridis,
height=4.5cm,
point meta max=41.9186855251517,
point meta min=-18.0813144748483,
tick align=outside,
tick pos=both,
width=0.9\columnwidth,
x grid style={white!69.01960784313725!black},
xlabel={Time frame index},
xmin=-0.5, xmax=1126.5,
xtick style={color=black},
y grid style={white!69.01960784313725!black},
ylabel={Frequency bin index},
ymin=-0.5, ymax=512.5,
ytick style={color=black}
]
\addplot graphics [includegraphics cmd=\pgfimage,xmin=-0.5, xmax=1126.5, ymin=-0.5, ymax=512.5] {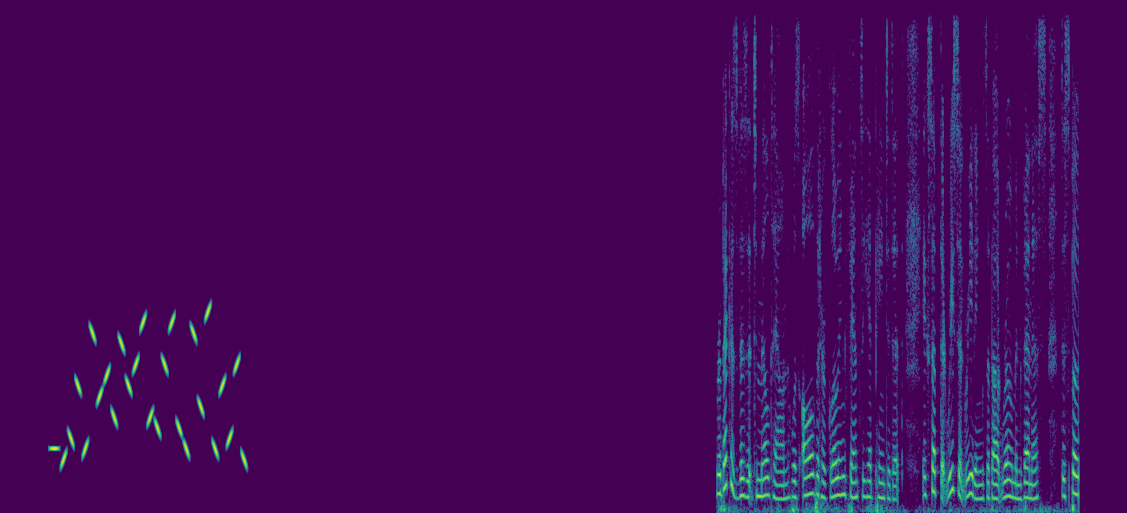};

\draw[red, line width=0.1em] (axis cs:40,30) rectangle (axis cs: 260,220);
\node[red, at={(axis cs:150,220)}, anchor=south, align=center, font=\footnotesize] {chirp\\sequence};

\end{axis}

\end{tikzpicture}

%% file: images/tikz_spect.tex
% This file was created by tikzplotlib v0.8.1.
\begin{tikzpicture}

\begin{groupplot}[group style={group size=3 by 1, horizontal sep=0.2cm}]
\nextgroupplot[
axis line style={white!80.0!black},
%colorbar,
%colorbar style={ylabel={Energy / dB}},
%colormap/viridis,
height=4.5cm,
point meta max=30.9552748697889,
point meta min=-29.0447251302111,
tick align=outside,
tick pos=both,
title={United Kingdom, Newmarket\\STOI=${0.36}{}$},
title style={align=center},
width=0.35\textwidth,
x grid style={white!80.0!black},
xlabel={Time frame index},
xmin=-0.5, xmax=182.5,
xminorgrids,
xtick style={color=white!15.0!black},
y grid style={white!80.0!black},
ylabel={Frequency bin index},
ymin=-0.5, ymax=512.5,
yminorgrids,
ytick style={color=white!15.0!black}
]
\addplot graphics [includegraphics cmd=\pgfimage,xmin=-0.5, xmax=182.5, ymin=-0.5, ymax=512.5] {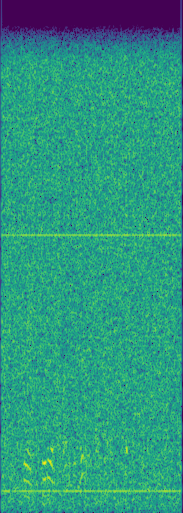};

\nextgroupplot[
axis line style={white!80.0!black},
%colorbar,
%colorbar style={ylabel={Energy / dB}},
%colormap/viridis,
height=4.5cm,
point meta max=36.2013570575531,
point meta min=-23.7986429424469,
tick align=outside,
tick pos=both,
title style={align=center},
title={Belgium, Manage\\STOI=0.48},
width=0.35\textwidth,
x grid style={white!80.0!black},
xlabel={Time frame index},
xmin=-0.5, xmax=182.5,
xminorgrids,
xtick style={color=white!15.0!black},
y grid style={white!80.0!black},
ymin=-0.5, ymax=512.5,
yminorgrids,
yticklabels={},
ytick style={color=white!15.0!black}
]
\addplot graphics [includegraphics cmd=\pgfimage,xmin=-0.5, xmax=182.5, ymin=-0.5, ymax=512.5] {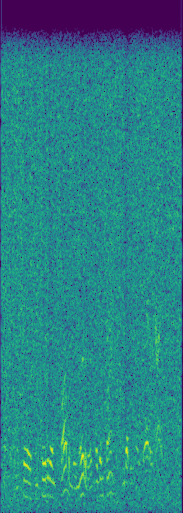};

\nextgroupplot[
axis line style={white!80.0!black},
colorbar,
colorbar style={ylabel={Energy / dB}, at={(1.05,1)}, width=0.3cm},
colormap/viridis,
height=4.5cm,
point meta max=40.8730387612487,
point meta min=-19.1269612387513,
tick align=outside,
tick pos=both,
title style={align=center},
title={Germany, Hagenow\\STOI=0.72},
width=0.35\textwidth,
x grid style={white!80.0!black},
xlabel={Time frame index},
xmin=-0.5, xmax=182.5,
xminorgrids,
xtick style={color=white!15.0!black},
y grid style={white!80.0!black},
ymin=-0.5, ymax=512.5,
yminorgrids,
yticklabels={},
ytick style={color=white!15.0!black}
]
\addplot graphics [includegraphics cmd=\pgfimage,xmin=-0.5, xmax=182.5, ymin=-0.5, ymax=512.5] {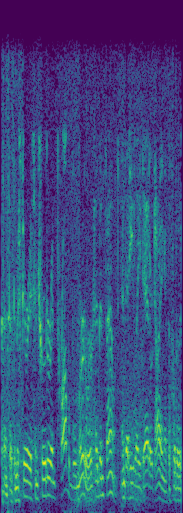};
\end{groupplot}

\end{tikzpicture}

%% file: sections/sigpro.tex
%% Tex magic
% !TeX spellcheck = en_US
% !TeX encoding = utf-8
% !TeX root = Heit2021.tex
% !TeX program = pdflatex
% !BIB program = bibtex

\section{Speech Activity Detection}\label{SEC:SAD}

Both the  \gls{DNN} based \gls{SAD}  and the statistical SAD are similar to the ones we proposed in \cite{Heitkaemper20Fearless}.
The following sections present a short overview over the systems.

\subsection{Statistical SAD}\label{SEC:SigPro}
The statistical \gls{SAD} processes the data in two phases. In the first phase the background noise is reduced by  Wiener filtering, whose noise \gls{PSD} is estimated by minimum statistics. The gain function of the Wiener filter is given by
\begin{align}
W(t,f) = \mathrm{max}\left(1-\gamma\frac{\overline{|V(t,f)|^{2}}}{|X(t,f)|^{2}}, G_{\mathrm{min}}\right).
\end{align}
with $G_{\mathrm{min}}$ denoting a lower bound on the Wiener filter gain.
Here, the \gls{STFT}-coefficients of the observed signal $X(t,f)$ with $t$ as the frame index and $f$ as the frequency bin index are used to determine the noise \gls{PSD} $\overline{|V(t,f)|^{2}}$ and the \gls{PSD} of the current analysis window $|X(t,f)|^{2}$.

Since the Kiwi-\gls{SDR}'s \gls{AGC} adapts to the receiver channel conditions, the noise level of the recordings is changing rapidly over time. Hence, the observation window of the minimum statistics for estimating $\overline{|V(t,f)|^{2}}$ has to be kept small, and the oversubtraction factor $\gamma$ has to be chosen quite large ($\gamma>20$).
Noise \gls{PSD} estimation and Wiener filtering is repeated several times to maximize the \gls{SNR} gain.
Following the multi-stage Wiener filter, a linear highpass is applied to remove low frequency noise.
Afterwards, a simple $1^{\text{st}}$-order \gls{LPC} filter suppresses all parts of the signals that are not as well predictable as the highly correlated speech signals.

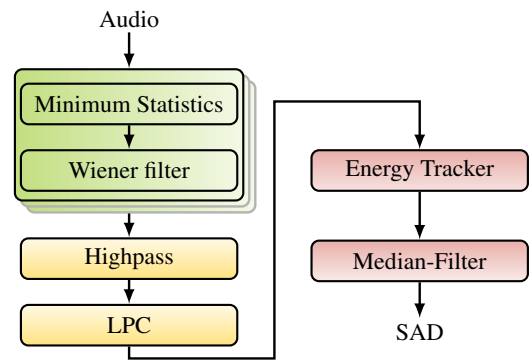
\begin{figure}[t]
	\centering
	\input{images/sigpro.tex}
	\caption{Overview on statistical \gls{SAD} components and signal processing queue.}
	\label{fig:sigpro}
	\vspace{-0.5cm}
\end{figure}

%In the second phase temporally smoothed sub-band energies are calculated, including adaptive thresholds to decide on speech activity.
%To this end, the \gls{STFT} of the denoised signal is computed and the frequency bins are smoothed with a mel filter-bank.
%Subsequently, sub-bands with $\SI{1}{kHz}$ bandwidth are formed, and the energy per sub-band is determined.
%Then, all sub-band energies are summed  with a weighing factor of $1/s$ for the $s$-th sub-band.
%The resulting value, called \gls{CSBE}$(t)$, is tracked with minimum statistics to estimate the noise floor level (F-CSBE).
%A frame is declared to contain speech if the \gls{CSBE}$(t)$ value exceeds the F-CSBE value by a certain factor.
%Fluctuations between speech and noise decisions are suppressed by a subsequent median filter.
In the second phase temporally smoothed sub-band energies and an adaptive threshold are calculated to decide on speech activity.
Figure \ref{fig:sigpro} depicts a block diagram of the statistical \gls{SAD} system.

%The strength of the statistical \gls{SAD} originates from the large \gls{SNR} gain of the multi-stage Wiener filter, which suppresses the noise without regard to speech quality. Thus, it focuses on parts of speech that have significant signal strength and misses parts with noise-like characteristics. However, the experimental results will show that these protected speech parts are sufficient for a noise robust \gls{SAD}.

%% file: images/sigpro.tex
\tikzset{%
  block/.style    = {draw, thick, rectangle, minimum height = 1.7em, minimum width = 9em, fill=white, align=center, rounded corners=0.1cm},
  sum/.style      = {draw, circle, node distance = 2cm}, % Adder
  cross/.style={path picture={\draw[black](path picture bounding box.south east) -- (path picture bounding box.north west)
		 (path picture bounding box.south west) -- (path picture bounding box.north east);}},
	   zigzag/.style = {% added for solution
	   	to path={ -- ($(\tikztostart)!.55!-9:(\tikztotarget)$) --
	   		($(\tikztostart)!.45!+9:(\tikztotarget)$) -- (\tikztotarget)
	   		\tikztonodes},sharp corners},
  multiple/.style = 	{double copy shadow={shadow xshift=0.25em,shadow yshift=-0.25em,draw=black!30, left color=lightgreen!50!white, right color=lightgreen!10!white},fill=white,draw=black,thick,minimum height = 5.5em,minimum width=9.5em, rounded corners=0.1cm,left color=lightgreen!50!white, right color=lightgreen!10!white},
   		}
\tikzstyle{branch}=[{circle,inner sep=0pt,minimum size=0.3em,fill=black}]
\tikzstyle{box} = [draw, dotted, inner xsep=4mm, inner ysep=3mm]
\tikzstyle{every path}=[line width=0.1em]
\begin{tikzpicture}[auto, line width=0.1em, node distance = 1em]

\node[multiple, align=center](shadow){};
\node[block, at={(shadow.north)}, fill=none, yshift=-1.5em](ms){Minimum Statistics};
\node[block, below= of ms,fill=none](wf){Wiener filter};
\node[block, below= 1.5em of shadow, bottom color=yellow!50!white, top color=yellow!10!white](hp){Highpass};
\node[block, below= of hp, bottom color=yellow!50!white, top color=yellow!10!white](lpc){LPC};
\node[block, right=3em of wf, top color=red!40!white, bottom color=red!15!white](et){Energy Tracker};
%\node[block, below= of et.south east, anchor=north east, minimum width=4em, top color=red!40!white, bottom color=red!15!white](em){EM-Alg.};
\node[block, below=2em of et, top color=red!40!white, bottom color=red!10!white](viterbi){Median-Filter};

\draw[-latex] (ms) -- (wf);
\draw[-latex] ($(shadow.south) - (0,0.5em)$) -- (hp);
\draw[-latex] (hp) -- (lpc);
\draw[-latex] (lpc.south) -- ($(lpc.south) - (0,0.5em)$) -| ($(wf) !0.5! (et)$) |- ($(et.north) + (0,2em)$) -- (et.north);
%\draw[-latex] ($(et.south) - (2.5em,0)$) -- ($(viterbi.north) - (2.5em,0)$);
\draw[-latex] (et.south) -- (viterbi.north);
%\draw[-latex] ($(et.south) - (2.5em,0)$) |- (em.west);
%\draw[-latex] (em.south) -- (em.south |- viterbi.north);
\draw[-latex] ($(shadow.north) + (0,1.5em)$) -- node[above, pos=0.1]{Audio} (shadow.north);
\draw[-latex] (viterbi.south) -- node[below, pos=0.9]{SAD} ($(viterbi.south) - (0,1.5em)$);
\end{tikzpicture}

%% file: sections/dnn.tex
%% Tex magic
% !TeX spellcheck = en_US
% !TeX encoding = utf-8
% !TeX root = ../Heit2021.tex
% !TeX program = pdflatex
% !BIB program = bibtex

\subsection{DNN based SAD} \label{SEC:DNN}
\begin{figure}[bh]
	\centering
	\input{images/model}
	\caption{Block diagram of the \gls{DNN} architecture for speech activity detection, where $R$ represents the output size of the FF layer.}
	\label{fig:model}
\end{figure}
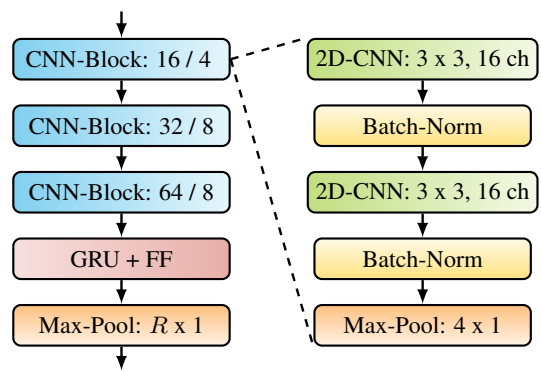
Since the Fearless Steps challenge database \cite{fearless20} also offers \gls{HF} transmission data, the best performing \gls{SAD} system \cite{Heitkaemper20Fearless} from the 2020 challenge is adapted to the new database.
%The requirement for a strong \gls{DNN} based \gls{SAD} are similar to the fearless database \cite{fearless20}. Therefore
%the architecture presented in  is adapted to the new database.
Fig. \ref{fig:model} shows the block diagram of the  architecture of the \gls{DNN} for \gls{SAD}.

The \gls{STFT} input data is first normalized using a $\ell_2$-norm over the time dimension to compensate for possible variations in the signal due to different recording devices, distance to the transmitter or speaker.
The normalized signal is processed by several \gls{CNN} blocks with subsequent temporal smoothing as shown in the figure.
The \gls{CNN} output is then processed by two uni-directional \gls{GRU} layers \cite{Kyunghyun14GRU} followed by a \gls{FF} classification layer and max pooling over the feature dimension.
All \gls{CNN} blocks consist of two 2D-\gls{CNN} layers with batch normalization and max pooling over the feature dimension.
Note, that no pooling is applied to the time dimension to allow a frame-wise activity estimation.
The purpose of the \gls{GRU} is to gather temporal information from a larger context than the \gls{CNN} layer.
During evaluation the network output is smoothed using a simple median filter over 25 frames.

%During training, each utterance is split into segments  of \SI{4}{s} length with a random offset to ensure that the network cannot overfit to the particular lengths of speech and noise regions of the training set.
%Training is performed using binary cross entropy, a batch size of 32 and the ADAM optimizer with a learning rate of $0.001$.
%During evaluation the network output is smoothed by applying a simple median filter with a window length of \SI{250}{ms} and \SI{50}{\percent} overlap.
%Then, the ``segmented \gls{RNN}'' operates on overlapping fixed-length segments of the input signal independently.
%It consists of two \gls{GRU} layers \cite{Kyunghyun14GRU} followed by a feed forward classification layer.
%The purpose of the GRU is to gather temporal information and conduct temporal smoothing, rendering a subsequent explicit smoothing of the network output, e.g., by a median filter, unnecessary.
%For each segment, only the last value of the  \gls{GRU} is kept. The outputs of all segments containing the same frame are combined to come up with a decision on speech presence or absence for that particular frame. Basing the decision only on the last output of the \gls{GRU} obviously leads to a reduction of time resolution, which, however, is considered acceptable. In this work the segmentation was chosen such that the time resolution is reduced by a factor of ten.

%% file: images/model.tex
\tikzset{%
  block/.style    = {draw, thick, rectangle, minimum height = 1.7em, minimum width = 9em, fill=white, align=center, rounded corners=0.1cm},
  sum/.style      = {draw, circle, node distance = 2cm}, % Adder
  cross/.style={path picture={\draw[black](path picture bounding box.south east) -- (path picture bounding box.north west)
		 (path picture bounding box.south west) -- (path picture bounding box.north east);}},
	   zigzag/.style = {% added for solution
	   	to path={ -- ($(\tikztostart)!.55!-9:(\tikztotarget)$) --
	   		($(\tikztostart)!.45!+9:(\tikztotarget)$) -- (\tikztotarget)
	   		\tikztonodes},sharp corners}
%wireless/.pic={
%	\draw [->] (0,0) -| (.5,#1);
%	\foreach \r in {.1,.2,.3}
%	\draw (.6,#1) ++ (60:\r) arc (60:-60:\r);
%},
%   connect/.style args={(#1) to (#2) over (#3) by #4}{
%         insert path={
%             let \p1=($(#1)-(#3)$), \n1={veclen(\x1,\y1)}, 
%             \n2={atan2(\y1,\x1)}, \n3={abs(#4)}, \n4={#4>0 ?180:-180}  in 
%             (#1) -- ($(#1)!\n1-\n3!(#3)$) 
%             arc (\n2:\n2+\n4:\n3) -- (#2)
%         }
%     },
               }
\tikzstyle{branch}=[{circle,inner sep=0pt,minimum size=0.3em,fill=black}]
\tikzstyle{box} = [draw, dotted, inner xsep=4mm, inner ysep=3mm]
\tikzstyle{every path}=[line width=0.1em]
\begin{tikzpicture}[auto, line width=0.1em, node distance = 1cm, font=\small]
\node[block, left color=lightblue!50!white, right color=lightblue!10!white, at={(0,0)}] (cnn2d) {CNN-Block: 16 / 4 } ;
\node[block, left color=lightblue!50!white, right color=lightblue!10!white, below= 1em of cnn2d] (cnn2d_2) {CNN-Block: 32 / 8} ;
\node[block, left color=lightblue!50!white, right color=lightblue!10!white, below= 1em of cnn2d_2] (cnn2d_3) {CNN-Block: 64 / 8} ;
\node[block, below= 1em of cnn2d_3, right color=red!40!white, left color=red!15!white] (smoothing) {GRU + FF} ;
\node[block, below= 1em of smoothing, top color=orange!50!white, bottom color=orange!10!white] (maxpool) {Max-Pool: $R$ x 1};

\node[block, right= 1 of cnn2d, left color=lightgreen!50!white, right color=lightgreen!10!white] (cnn2d_exp_1) {2D-CNN: 3 x 3, 16 ch} ;
\node[block, below= 1em of cnn2d_exp_1, bottom color=yellow!50!white, top color=yellow!10!white] (batch_norm) {Batch-Norm} ;
\node[block, below= 1em of batch_norm, left color=lightgreen!50!white, right color=lightgreen!10!white] (cnn2d_exp_2) {2D-CNN: 3 x 3, 16 ch} ;
\node[block, below= 1em of cnn2d_exp_2, bottom color=yellow!50!white, top color=yellow!10!white] (batch_norm_2) {Batch-Norm} ;
\node[block, below= 1em of batch_norm_2, top color=orange!50!white, bottom color=orange!10!white] (maxpool_exp) {Max-Pool: 4 x 1};

\draw[-latex] (0, 2em) -- (cnn2d);
\draw[-, dashed] (cnn2d.east) -- (cnn2d_exp_1.north west);
\draw[-, dashed] (cnn2d.east) -- (maxpool_exp.south west);
\draw[-latex] (cnn2d) -- (cnn2d_2);
\draw[-latex] (cnn2d_2) -- (cnn2d_3);
\draw[-latex] (cnn2d_3) -- (smoothing);
\draw[-latex] (smoothing) -- (maxpool);
\draw[-latex] (cnn2d_exp_1) -- (batch_norm);
\draw[-latex] (batch_norm) -- (cnn2d_exp_2);
\draw[-latex] (cnn2d_exp_2) -- (batch_norm_2);
\draw[-latex] (batch_norm_2) -- (maxpool_exp);
\draw[-latex] (maxpool.south) -- ($(maxpool.south) - (0,1em)$);

%\draw[-] (cnn2d) -- ($ (cnn2d) + (1,0) $) -- ($ (cnn2d) + (1.4,0.5) $);
%\draw[-latex] ($ (rnn) - (1.2,0) $) -- (rnn);
%\draw[-latex] ($ (rnn_seg) - (1.5,1) $) |- (rnn_seg);
%\draw[-latex] ($ (cnn1d) - (1.5,-1) $) |- (cnn1d);
%\draw[-] (rnn) -- ($ (rnn) + (1.2,0) $);
%\draw[->] ($ (maxpool) - (1.4,-0.5) $) -- ($ (maxpool) - (1.1,0) $) -- (maxpool);
%\draw[-] (rnn_seg) -| ($ (rnn_seg) + (1.5, -1) $);
%\draw[-] (cnn1d) -| ($ (cnn1d) + (1.5, 1) $);
%\draw[-latex] (maxpool) -- ($ (maxpool) + (1.3, 0) $);
\end{tikzpicture}

%% file: sections/experiments.tex
%% Tex magic
% !TeX spellcheck = en_US
% !TeX encoding = utf-8
% !TeX root = ../Heit2021.tex
% !TeX program = pdflatex
% !BIB program = bibtex

\section{Experiments} \label{SEC:Experiments}
\begin{table}[b]
	\renewcommand*{\arraystretch}{1.1}
	\centering
%	\small
	\begin{tabular}{
			l
			c
			c
			c
			c
			c
		}
		\toprule
		& {Training} & {Development} & {Evaluation}\\
		\toprule
		{Duration / h} & 121 & 18 & 37 \\
		{Speech / $\%$} & 13.90 & 12.58 & 13.16 \\
		{$\#~$Speakers} & 705 & 80 & 175 \\		
		Average SDR / dB & -8.80 & -8.31 & -7.91 \\
		Average STOI & 0.63 & 0.64 & 0.61 \\
		\toprule
	\end{tabular}
	\caption{%
		Statistics of the different data sets.
		\label{tbl:datasets} 
	}
\end{table}
The recorded database is split into the following three data sets: training, development and evaluation.
Each set includes data recorded by about $22$ different KiwiSDR stations.
The speakers are strictly disjoint among the three sets and the speakers are uniformly distributed between female and male over the whole database.
No speaker is active in more than one example.

To classify the degree of distortion of each example we measured the \gls{SDR} \cite{Vincent2006Performance} and the short time objective intelligibility (STOI) score~\cite{Taal2011STOI}. Both metrics require the availability of clean and distorted data as provided by this database.

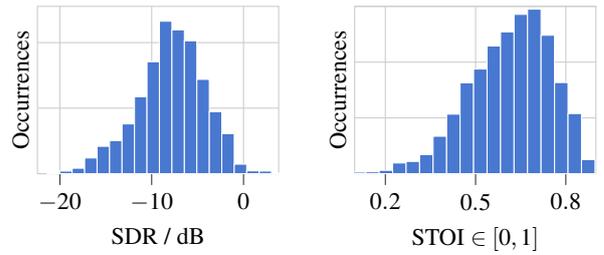
\begin{figure}[t]
	\centering
	\input{images/density}
	\caption{Histogram of STOI and SDR values on evaluation data.}
	\label{fig:hist}
\end{figure}

In Fig \ref{fig:hist} the histograms of SDR and STOI values in the evaluation set are presented, illustrating the diversity of channel conditions represented by the database.
The low \gls{SDR} values can be explained by the \gls{AGC} which significantly changes the signal amplitudes compared to the original clean signal. Thereby, leading to very low \gls{SDR} values even in case of low noise energy levels. This is a challenge for many current neural network based enhancement systems which rely on SI-\gls{SDR} loss functions \cite{Luo2018TasNetSI}.

Table \ref{tbl:datasets} summarizes the duration, the percentage of speech activity, the number of speakers, as well as the average values of SDR and STOI, for the training, development and evaluation set, respectively.
For all data sets the percentage of speech activity is close to $\SI{13}{\percent}$.
Therefore, the database includes a large amount of ham radio noise which could be used for data augmentation.

To evaluate the \gls{SAD} systems, we used the OpenSAD challenge scoring \cite{nist16} with a collar of \SI{0.5}{s} around both the beginning and end of each segment, i.e., speech on/offset errors below that value were not counted as an error.

We use precision ($P$), recall ($R$) and $F_1$-score$ = 2 (P\cdot R)/(P+R)$ as metrics with\\
\begin{minipage}{0.5\columnwidth}
	\begin{align}
	P &=\frac{\mathrm{TP}}{\mathrm{TP} + \mathrm{FP}},
	\end{align}
\end{minipage}
\begin{minipage}{0.49\columnwidth}
	\begin{align}
	R &= \frac{\mathrm{TP}}{\mathrm{TP} + \mathrm{FN}},
	\end{align}
\end{minipage}\\
\\
where TP, FP, TN and FN are the number of true positive, false positive, true negative and false negative predictions.
For the experiments the following \gls{STFT} parameters are used:
\begin{itemize}
	\item DNN SAD: FFT size 256, \SI{10}{ms} shift, \SI{25}{ms} window length 
	\item Stat. SAD: FFT size 1024, \SI{32}{ms} shift, \SI{64}{ms} window length
\end{itemize}
%\vspace*{-2em}
Furthermore, the number of hidden units is set to $514$ for the \gls{GRU} and to $10$ for the \gls{FF} classifier layer.

\begin{figure}[t]
	\centering
	\input{images/comparison_nn_sigpro}
	\caption{ROC curves of the \gls{SAD} systems for the development data set. At the intersection with the straight gray line the equal error rate can be read off.}
	\label{fig:comparison}
\end{figure}
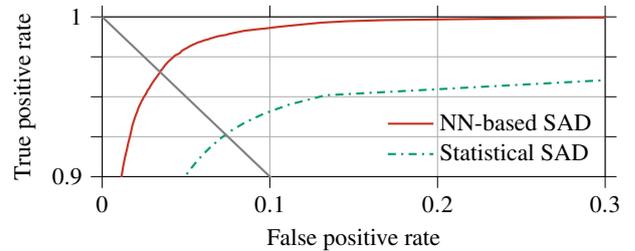

The results for the \gls{DNN} and the statistical \gls{SAD} on the development data set  are displayed in terms of a \gls{ROC} curve in Figure \ref{fig:comparison}.
From the \gls{ROC} curve the threshold corresponding to the equal error rate is determined for each system on the development data set and then applied on the evaluation data set.
The results are shown in Table \ref{tbl:comparison}.  

The \gls{DNN} outperforms the statistical \gls{SAD} on both the development and evaluation data set.
However, the statistical \gls{SAD} using a C implementation has a significantly lower \gls{RT} factor than the \gls{DNN} implemented in Pytorch \cite{pytorch19}.
The real-time (RT) factor is calculated on an Intel\textsuperscript \textregistered Xeon\textsuperscript \textregistered CPU E3-1240 v6 @ 3.70GHz with 8GB RAM.
For both systems all detection metrics are lower compared to the values obtained on the 2020 Fearless challenge \cite{Heitkaemper20Fearless} which emphasizes the difficulty of the presented database.

%The automatic smoothing outperforms the statistic one for both features and both features achieve similar results for both.
%However, the magnitude of the \gls{STFT} coefficient slightly outperforms the Mel feature banks for the automatic smoothing with a \gls{EER} of \SI{2.43}{\percent}.

\begin{table}[h]
	\renewcommand*{\arraystretch}{1.1}
	\centering
%	\small
	\begin{tabular}{
			l
			S[table-auto-round, table-format=-1.2]
			S[table-auto-round, table-format=-1.2]
			S[table-auto-round, table-format=-1.2]
			S[table-auto-round, table-format=-1.4]
		}
		\toprule
		System & {$R$} & {$P$} & {F1-score} & {RT-factor}\\
		\toprule
		Statistical & 85.90045356 & 93.23647535 & 89.41825194 & 0.00472 \\
%		\gls{RNN} & 99.20801416 & 80.7376812 & 89.02491392 \\
		DNN based & 95.58814162 & 95.65484254 & 95.62148045 & 0.0119\\		
		\toprule
	\end{tabular}

\caption{%
	Results of different \gls{SAD} systems on the evaluation data set, where the SAD threshold has been determined on the development set.
	\label{tbl:comparison}
}
\end{table}

%However, the magnitude of the \gls{STFT} coefficient slightly outperforms the Mel feature banks for the automatic smoothing with a \gls{EER} of \SI{2.43}{\percent}.

%The presented database is evaluated in terms of the range of \gls{SNR}, \gls{STOI}, \gls{PESQ} witnessed on the 

%\begin{figure}[htb]
%	\centering
%	\def\svgwidth{1.0\columnwidth}
%	\import{images/}{ErrorCalc.pdf_tex}
%	\caption{System for distributed recording of radio signals.}
%	\label{FIG:ErrorCalc}
%\end{figure}
%
%\begin{figure}[htb]
%\includegraphics[width=0.75\columnwidth]{images/paramsweep.eps}	
%\end{figure}
%
%\begin{itemize}
%\item Variation der Entscheidungsparamter
%\begin{itemize}
%	\item Energiebasierte SAD: $[0.01:0.1:5, 5.5:0.5:10, 15:5:50]$
%	\item BIC Gradienten: $[0.01:0.05:1]$
%\end{itemize}
%\end{itemize}
%
%\begin{figure}[htb]
%	\centering
%	\input{images/cnn_vss_rnn}
%\end{figure}
%

%
%
%\begin{figure}[htb]
%\input{images/comparison_tikz}	
%\end{figure}

%% file: images/density.tex
% This file was created by tikzplotlib v0.8.4.
\begin{tikzpicture}

\definecolor{color0}{rgb}{0.282352941176471,0.470588235294118,0.815686274509804}

\begin{groupplot}[group style={group size=2 by 1}]
\nextgroupplot[
axis line style={white!80.0!black},
height=3.8cm,
tick align=outside,
width=0.58\columnwidth,
x grid style={white!80.0!black},
xlabel={SDR /\ dB},
xmajorgrids,
xtick={-20, -10 , 0},
xtick pos=bottom,
xmin=-22.4469371234811, xmax=3.82822769784994,
xminorgrids,
xtick style={color=white!15.0!black},
y grid style={white!80.0!black},
ylabel={Occurrences},
ymajorgrids,
ymajorticks=false,
ymin=0, ymax=0.127669523450976,
yminorgrids,
ytick style={color=white!15.0!black}
]
\draw[draw=white,fill=color0] (axis cs:-21.3791639758279,0) rectangle (axis cs:-20.0212085143217,0.000363565132992945);
\draw[draw=white,fill=color0] (axis cs:-20.0212085143216,0) rectangle (axis cs:-18.6632530528155,0.0019996082314612);
\draw[draw=white,fill=color0] (axis cs:-18.6632530528155,0) rectangle (axis cs:-17.3052975913093,0.00418099902941888);
\draw[draw=white,fill=color0] (axis cs:-17.3052975913093,0) rectangle (axis cs:-15.9473421298031,0.0119976493887672);
\draw[draw=white,fill=color0] (axis cs:-15.9473421298031,0) rectangle (axis cs:-14.5893866682969,0.0209049951470944);
\draw[draw=white,fill=color0] (axis cs:-14.5893866682969,0) rectangle (axis cs:-13.2314312067907,0.0252677767430097);
\draw[draw=white,fill=color0] (axis cs:-13.2314312067907,0) rectangle (axis cs:-11.8734757452845,0.0379925563977628);
\draw[draw=white,fill=color0] (axis cs:-11.8734757452845,0) rectangle (axis cs:-10.5155202837783,0.0587157689783607);
\draw[draw=white,fill=color0] (axis cs:-10.5155202837783,0) rectangle (axis cs:-9.15756482227208,0.0854378062533422);
\draw[draw=white,fill=color0] (axis cs:-9.15756482227208,0) rectangle (axis cs:-7.79960936076588,0.116340842557743);
\draw[draw=white,fill=color0] (axis cs:-7.79960936076588,0) rectangle (axis cs:-6.44165389925969,0.10979667016387);
\draw[draw=white,fill=color0] (axis cs:-6.44165389925969,0) rectangle (axis cs:-5.08369843775349,0.101434672105032);
\draw[draw=white,fill=color0] (axis cs:-5.08369843775349,0) rectangle (axis cs:-3.72574297624729,0.0719858963326032);
\draw[draw=white,fill=color0] (axis cs:-3.72574297624729,0) rectangle (axis cs:-2.3677875147411,0.0472634672890829);
\draw[draw=white,fill=color0] (axis cs:-2.3677875147411,0) rectangle (axis cs:-1.0098320532349,0.0305394711714074);
\draw[draw=white,fill=color0] (axis cs:-1.0098320532349,0) rectangle (axis cs:0.348123408271299,0.00727130265985891);
\draw[draw=white,fill=color0] (axis cs:0.348123408271299,0) rectangle (axis cs:1.7060788697775,0.00218139079795767);
\draw[draw=white,fill=color0] (axis cs:1.7060788697775,0) rectangle (axis cs:3.06403433128369,0.0019996082314612);
\draw[draw=white,fill=color0] (axis cs:3.06403433128369,0) rectangle (axis cs:4.42198979278989,0.000363565132992945);
\draw[draw=white,fill=color0] (axis cs:4.42198979278989,0) rectangle (axis cs:5.77994525429609,0.000363565132992946);

\nextgroupplot[
axis line style={white!80.0!black},
height=3.8cm,
tick align=outside,
width=0.58\columnwidth,
x grid style={white!80.0!black},
xlabel={STOI $\in[0,1]$},
xmajorgrids,
xmin=0.0980789618669371, xmax=0.843455472110445,
xtick={0.2,0.5,0.8},
xtick pos=bottom,
xmax=0.9,
xminorgrids,
xtick style={color=white!15.0!black},
y grid style={white!80.0!black},
ylabel={Occurrences},
ymajorgrids,
ymajorticks=false,
ymin=0, ymax=3,
yminorgrids,
ytick style={color=white!15.0!black}
]
\draw[draw=white,fill=color0] (axis cs:9.99999999999959e-06,0) rectangle (axis cs:0.0448596449615906,0.0825600612419243);
\draw[draw=white,fill=color0] (axis cs:0.0448596449615906,0) rectangle (axis cs:0.0897092899231811,0.00550400408279496);
\draw[draw=white,fill=color0] (axis cs:0.0897092899231811,0) rectangle (axis cs:0.134558934884772,0.0275200204139748);
\draw[draw=white,fill=color0] (axis cs:0.134558934884772,0) rectangle (axis cs:0.179408579846362,0.0330240244967697);
\draw[draw=white,fill=color0] (axis cs:0.179408579846362,0) rectangle (axis cs:0.224258224807953,0.0605440449107445);
\draw[draw=white,fill=color0] (axis cs:0.224258224807953,0) rectangle (axis cs:0.269107869769543,0.192640142897823);
\draw[draw=white,fill=color0] (axis cs:0.269107869769543,0) rectangle (axis cs:0.313957514731134,0.220160163311798);
\draw[draw=white,fill=color0] (axis cs:0.313957514731134,0) rectangle (axis cs:0.358807159692725,0.341248253133287);
\draw[draw=white,fill=color0] (axis cs:0.358807159692725,0) rectangle (axis cs:0.403656804654315,0.671488498100985);
\draw[draw=white,fill=color0] (axis cs:0.403656804654315,0) rectangle (axis cs:0.448506449615906,1.06777679206222);
\draw[draw=white,fill=color0] (axis cs:0.448506449615906,0) rectangle (axis cs:0.493356094577496,1.62918520850731);
\draw[draw=white,fill=color0] (axis cs:0.493356094577496,0) rectangle (axis cs:0.538205739539087,1.87686539223308);
\draw[draw=white,fill=color0] (axis cs:0.538205739539087,0) rectangle (axis cs:0.583055384500677,2.2896656984427);
\draw[draw=white,fill=color0] (axis cs:0.583055384500677,0) rectangle (axis cs:0.627905029462268,2.54284988625127);
\draw[draw=white,fill=color0] (axis cs:0.627905029462268,0) rectangle (axis cs:0.672754674423858,2.85657811897059);
\draw[draw=white,fill=color0] (axis cs:0.672754674423858,0) rectangle (axis cs:0.717604319385449,2.95014618837809);
\draw[draw=white,fill=color0] (axis cs:0.717604319385449,0) rectangle (axis cs:0.76245396434704,2.48230584134052);
\draw[draw=white,fill=color0] (axis cs:0.76245396434704,0) rectangle (axis cs:0.80730360930863,1.6346892125901);
\draw[draw=white,fill=color0] (axis cs:0.80730360930863,0) rectangle (axis cs:0.852153254270221,1.07878480022781);
\draw[draw=white,fill=color0] (axis cs:0.852153254270221,0) rectangle (axis cs:0.897002899231811,0.253184187808568);
\end{groupplot}

\end{tikzpicture}

%% file: images/comparison_nn_sigpro.tex
% This file was created by tikzplotlib v0.8.1.
\begin{tikzpicture}

\begin{axis}[
height=3.7cm,
width=\columnwidth,
legend cell align={left},
legend style={fill=none,at={(0.97,0.01)}, anchor=south east, draw=none, xshift=0.2cm},
tick align=outside,
tick pos=both,
x grid style={white!69.01960784313725!black},
xlabel={False positive rate},
xmajorgrids,
xmin=0, xmax=0.3,
xtick={0,0.1,0.2,0.3},
xtick style={color=black},
y grid style={white!69.01960784313725!black},
ylabel={True positive rate},
ymajorgrids,
ymin=0.9, ymax=1,
ytick={0.9, 0.925, 0.95, 0.975, 1},
yticklabels={0.9, , , , 1},
ytick style={color=black}
]
\addplot [thick, red]
table {%
1 1
1 1
0.29568802480343 0.999557221020603
0.207325982065522 0.998500394090476
0.207325982065522 0.998500394090476
0.207325982065522 0.998500394090476
0.16806099832077 0.997953937238801
0.145614504482561 0.997265997323851
0.145614504482561 0.997265997323851
0.145614504482561 0.997265997323851
0.130538032590217 0.996379293753208
0.119480177732752 0.995261749395117
0.119480177732752 0.995261749395117
0.119480177732752 0.995261749395117
0.110677140056873 0.9943567160349
0.103418143760491 0.993558797382506
0.103418143760491 0.993558797382506
0.103418143760491 0.993558797382506
0.097555301620269 0.992761451536036
0.0926617954126212 0.992279721753794
0.0926617954126212 0.992279721753794
0.0926617954126212 0.992279721753794
0.088509929961246 0.991752167497617
0.0847434355092656 0.991301369235281
0.0847434355092656 0.991301369235281
0.0847434355092656 0.991301369235281
0.081565403455217 0.990592235501136
0.0788679828306197 0.990037759366522
0.0788679828306197 0.990037759366522
0.0788679828306197 0.990037759366522
0.0761397854374208 0.989455788547547
0.0737987227552845 0.988712286457952
0.0737987227552845 0.988712286457952
0.0737987227552845 0.988712286457952
0.0715389671810297 0.988345690666471
0.0695136088874136 0.987960765085417
0.0695136088874136 0.987960765085417
0.0695136088874136 0.987960765085417
0.0676963685705875 0.987465287960994
0.0659852155105742 0.986918831109319
0.0659852155105742 0.986918831109319
0.0659852155105742 0.986918831109319
0.0645192183723728 0.986495527531344
0.0630427271497485 0.985852839284405
0.0630427271497485 0.985852839284405
0.0630427271497485 0.985852839284405
0.0616923413113655 0.985544096891268
0.0603178808087182 0.985150006415426
0.0603178808087182 0.985150006415426
0.0603178808087182 0.985150006415426
0.0591658243473694 0.984647082813989
0.0579232012918835 0.984254710755921
0.0579232012918835 0.984254710755921
0.0579232012918835 0.984254710755921
0.0568067718062224 0.983897852665151
0.0559170674554456 0.983373735244519
0.0559170674554456 0.983373735244519
0.054943939542786 0.982964179008725
0.054943939542786 0.982964179008725
0.0540696676691016 0.982587845516533
0.0540696676691016 0.982587845516533
0.0540696676691016 0.982587845516533
0.0530185837007292 0.981999573832392
0.0520879612392614 0.98156424133001
0.0520879612392614 0.98156424133001
0.0520879612392614 0.98156424133001
0.0512087509729074 0.980960503885915
0.0503542326699014 0.980412901422392
0.0503542326699014 0.980412901422392
0.0503542326699014 0.980412901422392
0.0494444220061127 0.97999818993328
0.0486641559643173 0.979471208483027
0.0486641559643173 0.979471208483027
0.0486641559643173 0.979471208483027
0.0477963216382201 0.978998643595572
0.0471468348165855 0.978305548427304
0.0471468348165855 0.978305548427304
0.0471468348165855 0.978305548427304
0.0464563416986766 0.97748987279126
0.0457498869901749 0.976992104443141
0.0457498869901749 0.976992104443141
0.0457498869901749 0.976992104443141
0.0449757939392165 0.976608897279859
0.0442864472338916 0.976213088386245
0.0442864472338916 0.976213088386245
0.0442864472338916 0.976213088386245
0.0435621790375461 0.975947879243346
0.0429146322987459 0.97536705403622
0.0429146322987459 0.97536705403622
0.0429146322987459 0.97536705403622
0.0423664707124213 0.974779355158003
0.0417400885136338 0.974176763325757
0.0417400885136338 0.974176763325757
0.0417400885136338 0.974176763325757
0.041068378924986 0.973651500293277
0.04046157392571 0.972906852591832
0.04046157392571 0.972906852591832
0.04046157392571 0.972906852591832
0.0399236418670581 0.972461782388738
0.0393270663954548 0.971872365092749
0.0393270663954548 0.971872365092749
0.0393270663954548 0.971872365092749
0.0387158521170096 0.971164949776377
0.0381038441683138 0.970484456338441
0.0381038441683138 0.970484456338441
0.0381038441683138 0.970484456338441
0.0375548889117388 0.969802817288658
0.0370484391063555 0.969106858090769
0.0370484391063555 0.969106858090769
0.0365975462185052 0.968527178495491
0.0365975462185052 0.968527178495491
0.0360971812183755 0.967662814355891
0.0360971812183755 0.967662814355891
0.0355416120430466 0.966972583217245
0.0355416120430466 0.966972583217245
0.0349975069935577 0.966478824510595
0.0349975069935577 0.966478824510595
0.0349975069935577 0.966478824510595
0.0345337390105331 0.965560616614121
0.0341600085081451 0.964823415389691
0.0341600085081451 0.964823415389691
0.0341600085081451 0.964823415389691
0.0336926249162016 0.964248318241807
0.0331680970662244 0.963428632964294
0.0331680970662244 0.963428632964294
0.0331680970662244 0.963428632964294
0.0327744366219913 0.962860409487499
0.0323290994258942 0.962207983539849
0.0323290994258942 0.962207983539849
0.0323290994258942 0.962207983539849
0.0317588914437232 0.961302377373708
0.0311677834782899 0.960645368978664
0.0311677834782899 0.960645368978664
0.0311677834782899 0.960645368978664
0.0306717395717461 0.959752364542855
0.0301775475624535 0.959034065913923
0.0301775475624535 0.959034065913923
0.0301775475624535 0.959034065913923
0.0296382045344674 0.958186313146125
0.0290965686813134 0.957265814025955
0.0290965686813134 0.957265814025955
0.0290965686813134 0.957265814025955
0.0285940872271825 0.956351042965027
0.0279725552354811 0.955176218014517
0.0279725552354811 0.955176218014517
0.0279725552354811 0.955176218014517
0.0274604615527612 0.95455014113938
0.0269163565032723 0.953649690226556
0.0269163565032723 0.953649690226556
0.0269163565032723 0.953649690226556
0.0264426235493251 0.952495486289317
0.0259665977702099 0.951538327589999
0.0259665977702099 0.951538327589999
0.0259665977702099 0.951538327589999
0.0253236366817458 0.95061381882836
0.0247235337868071 0.949637184727619
0.0247235337868071 0.949637184727619
0.0247235337868071 0.949637184727619
0.0240785444299252 0.948289945193929
0.0233638884621686 0.946982802074932
0.0233638884621686 0.946982802074932
0.0233638884621686 0.946982802074932
0.0227165180945353 0.945470021629152
0.0219743482247625 0.943894805337635
0.0219743482247625 0.943894805337635
0.0219743482247625 0.943894805337635
0.0213941752716689 0.94239920906958
0.0208082702556553 0.940512959161229
0.0208082702556553 0.940512959161229
0.0208082702556553 0.940512959161229
0.0200982881238182 0.938680553009751
0.0194505650138513 0.936789720654007
0.0194505650138513 0.936789720654007
0.0194505650138513 0.936789720654007
0.0186919044399854 0.93432550956815
0.0179135784810245 0.931728407507882
0.0179135784810245 0.931728407507882
0.0179135784810245 0.931728407507882
0.0172809351058165 0.928400977894274
0.0164956424857682 0.924836406628052
0.0164956424857682 0.924836406628052
0.0164956424857682 0.924836406628052
0.0156270144894205 0.921496948090036
0.01477055610358 0.9172547474155
0.01477055610358 0.9172547474155
0.01477055610358 0.9172547474155
0.0137660341232351 0.912891111885036
0.012597222401043 0.906984337194809
0.012597222401043 0.906984337194809
0.012597222401043 0.906984337194809
0.0113738238027352 0.899171837194809
0.00974715253160527 0.888214632671017
0.00974715253160527 0.888214632671017
0.00974715253160527 0.888214632671017
0.00787338525582902 0.873573713248772
0.00578462152775879 0.843461305814209
0.00578462152775879 0.843461305814209
0.00578462152775879 0.843461305814209
0.00314496246027901 0.751951549783708
0 0
0 0
};
\addlegendentry{NN-based SAD}
\addplot [thick, green, dash dot]
table {%
1 1
0.133219305101006 0.950945749324465
0.111139369547604 0.944994042118824
0.0959847841830902 0.939214177081811
0.0850651535835218 0.933840250580868
0.0766597918288406 0.92862530381078
0.0698524266878987 0.923350675329433
0.0642341432717536 0.918140359036943
0.0595350280598716 0.913239770245992
0.0554732878251712 0.908335579972464
0.051967805673611 0.903809030871909
0.0489383490699051 0.899396700218764
0.0462336753291138 0.895055370222119
0.0438658040710813 0.890749026056212
0.0417168617760256 0.886556900337716
0.0397758381891032 0.882457898668686
0.0380553402169538 0.878351694034504
0.0365150956780186 0.874422991041569
0.0350842127212006 0.870481425610863
0.0337495975088942 0.86646422904606
0.0325270834719129 0.862779397873273
0.03141095207982 0.859071928810008
0.0303599156142364 0.855416423995341
0.0293468666523184 0.851846325767476
0.0284257839380812 0.848403822922305
0.0275478417887965 0.844819833261647
0.0267220256984669 0.841266198954129
0.0259441476711154 0.837838102039262
0.0252299127742662 0.834501071183818
0.0245130045095643 0.831346686944728
0.0238594863355934 0.828129019511804
0.0232422710315338 0.824805880089152
0.022652641316192 0.821641720397357
0.0220869330161886 0.818417878994301
0.021560159554734 0.815241371362245
0.0210599882172699 0.812191430117861
0.0205831899977185 0.809088495629858
0.0201260948588467 0.806156374315461
0.0196950113508886 0.803165600284826
0.0192618735776412 0.800338436462644
0.0188573412898384 0.797454163416757
0.012987906274772 0.743081551971452
0.00990930421398719 0.699894116412265
0.007991507862012 0.662989209958202
0.00661493100215562 0.630066514238204
0.00560278158013527 0.599845444947737
0.00481728106134122 0.57283844160762
0.00421639184630336 0.547706267197079
0.00374663730969068 0.524808555476209
0.00336617054313797 0.503954942365989
0.0030341988818447 0.484405580445802
0.00275367502086256 0.466441899854089
0.00250866694445276 0.449325082165253
0.00229371077059809 0.43367406788486
0.00209785466356094 0.41911841880509
0.00193668759629056 0.405791904278767
0.00179724993570495 0.393099250685825
0.00168728100502878 0.380941674504179
0.00157763232653193 0.369274414450383
0 0
};
\addlegendentry{Statistical SAD}
\addplot [thick, gray, forget plot]
table {%
0 1
1 0
};
\end{axis}

\end{tikzpicture}